%% file: main.tex
\documentclass[useAMS,twocolumn,usenatbib]{mn2e}

\usepackage[english]{babel}
\usepackage{graphicx,enumitem, booktabs}
\usepackage{blindtext}
\usepackage{amssymb,amsmath, wasysym}


\begin{document}

\title[The Lyman-$\alpha$ forest in $f(R)$ modified gravity]
{The Lyman-$\alpha$ forest in $f(R)$ modified gravity}
\author[C. Arnold, E. Puchwein \& V. Springel]
{Christian Arnold$^{1,2}$, 
Ewald Puchwein$^{3,1}$ and
Volker Springel$^{1,4}$
\\$^1$Heidelberger Institut f{\"u}r Theoretische Studien, Schloss-Wolfsbrunnenweg 35, 69118 Heidelberg, Germany
\\$^2$Institut f{\"u}r Theoretische Physik, Philosophenweg 16, 69120 Heidelberg, Germany
\\$^3$Institute of Astronomy and Kavli Institute for Cosmology, University of Cambridge, Madingley Road, Cambridge CB3 0HA, UK
\\$^4$Zentrum f\"ur Astronomie der Universit\"at Heidelberg, Astronomisches Recheninstitut, M\"{o}nchhofstr. 12-14, 69120 Heidelberg, Germany}
\date{\today}
\maketitle

\begin{abstract} 
  In this work, we analyze the Lyman-$\alpha$
  forest in cosmological hydrodynamical
  simulations of chameleon-type $f(R)$ gravity with the goal to assess
  whether the impact of such models is detectable in absorption line
  statistics. We carry out a set of hydrodynamical
  simulations with the cosmological simulation code
  \textsc{mg-gadget}, including star formation and cooling effects,
  and create synthetic Lyman-$\alpha$ absorption spectra from the
  simulation outputs. We statistically compare simulations with $f(R)$
  and ordinary general relativity, focusing on flux probability distribution functions (PDFs)
  and flux power-spectra, an analysis of the column
  density and line width distributions, as well as the matter
  power spectrum. We find that the influence of $f(R)$
  gravity on the Lyman-$\alpha$ forest is rather small. Even
  models with strong modifications of gravity, like $|\bar{f}_{R0}| = 10^{-4}$, do not change the
  statistical Lyman-$\alpha$ properties by more than $10\%$. The
  column density and line width distributions are hardly affected at
  all. It is therefore not possible to get competitive constraints on the background field
  $f_R$ using current observational data. An improved understanding of systematics in the observations and a more accurate modeling of the baryonic/radiative physics would be required to allow this in the future. The impact of $f(R)$ on
  the matter power spectrum in our results is consistent with previous
  works.
\end{abstract}

\section{Introduction}
\label{sec:introduction}
Theoretically explaining the late time accelerated expansion of the
Universe is one of the biggest challenges in modern cosmology. In the
standard model of cosmology, the \textit{$\Lambda$ cold dark matter}
($\Lambda$CDM) model, the cosmological constant $\Lambda$ is
responsible for the acceleration. While the $\Lambda$CDM model is
quite successful in explaining many aspects of cosmic structure
formation, it also features some problems, one of them being the lack
of a natural explanation for $\Lambda$. This encourages the search for
alternative theories.

In general, the models attempting to provide a theoretical explanation
for the accelerated expansion can be divided into two main classes.
The first class, which also includes $\Lambda$CDM, contains so-called
\textit{dark
  energy} models. These models add a new type of ``matter'' to the
energy-momentum tensor of general relativity (GR). If this type of
matter is described by an equation of state with a negative effective
pressure, it can drive the accelerated expansion.

The second class are the so-called \textit{modified gravity} models.
To explain the accelerated expansion, they modify the gravitational
interaction. In other words: while the dark energy models change the
right hand side of Einstein's equation, the modified gravity models
modify its left hand side.
 
In this work we consider $f(R)$-gravity, which can be viewed as a
member of the second class. The modification to gravity is carried out
by adding a scalar function of the Ricci scalar $R$ to the action of
GR. Because GR is well tested in our local environment, particularly
in the solar system, modified gravity theories typically share the
need for some screening mechanism which suppresses the modification
of GR in high density environments. Examples for such screening
mechanisms are the \textit{Chameleon} \citep{khoury2004}, the
\textit{Symmetron} \citep{hinterbichler2010}, the \textit{Vainshtein}
\citep{vainshtein1972} or the \textit{Dilaton} \citep{gasperini2002}
mechanism. In $f(R)$ gravity, the screening mechanism is determined by
the particular choice of $f(R)$. We here consider a model proposed by
\cite{husa2007}, which gives rise to a Chameleon mechanism.

\begin{figure*}
  \centerline{\includegraphics[width=\linewidth]{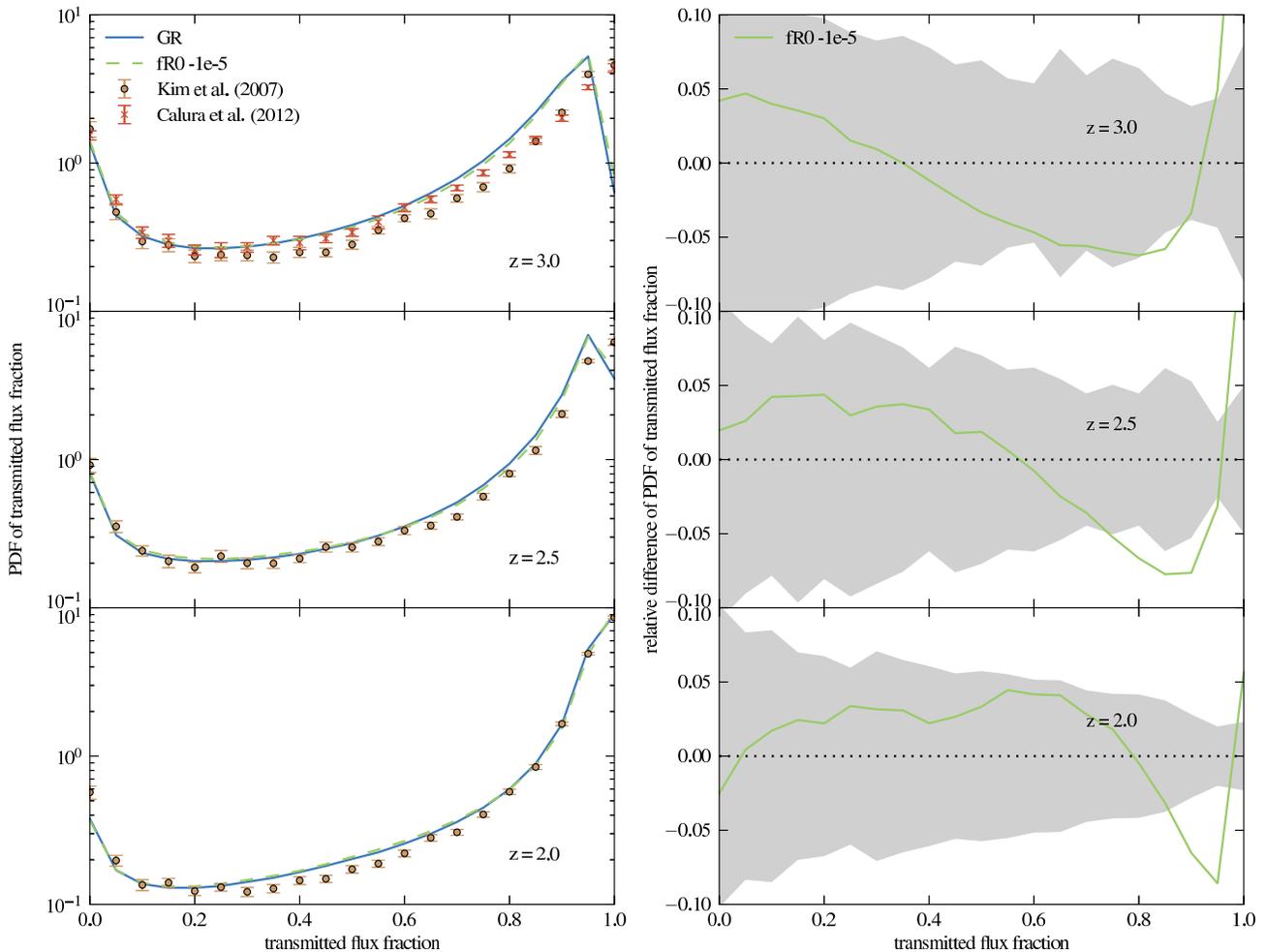}}
  \caption{\textit{Left panel:} PDF of the transmitted flux fraction
    for different redshifts for $\Lambda$CDM and $|\bar{f}_{R0}| =
    10^{-5}$, using the results of the large simulation boxes. The dots
    with errorbars show the data of \protect\cite{kim2007}. For $z = 3$, the observational results of \protect\cite{calura2012} are shown in addition (we plot the ``no metals, no LLS'' values of this work here).
    \textit{Right panel:} relative difference of the PDFs on the left
    hand side. The shaded regions show the $1 \sigma$ relative errors
    of the observational results of \protect\cite{kim2007}. The mean
    transmission is tuned to the values of this work in both panels.}
\label{fig:fluxpdf}
\end{figure*}

\begin{figure*}
  \centerline{\includegraphics[width=\linewidth]{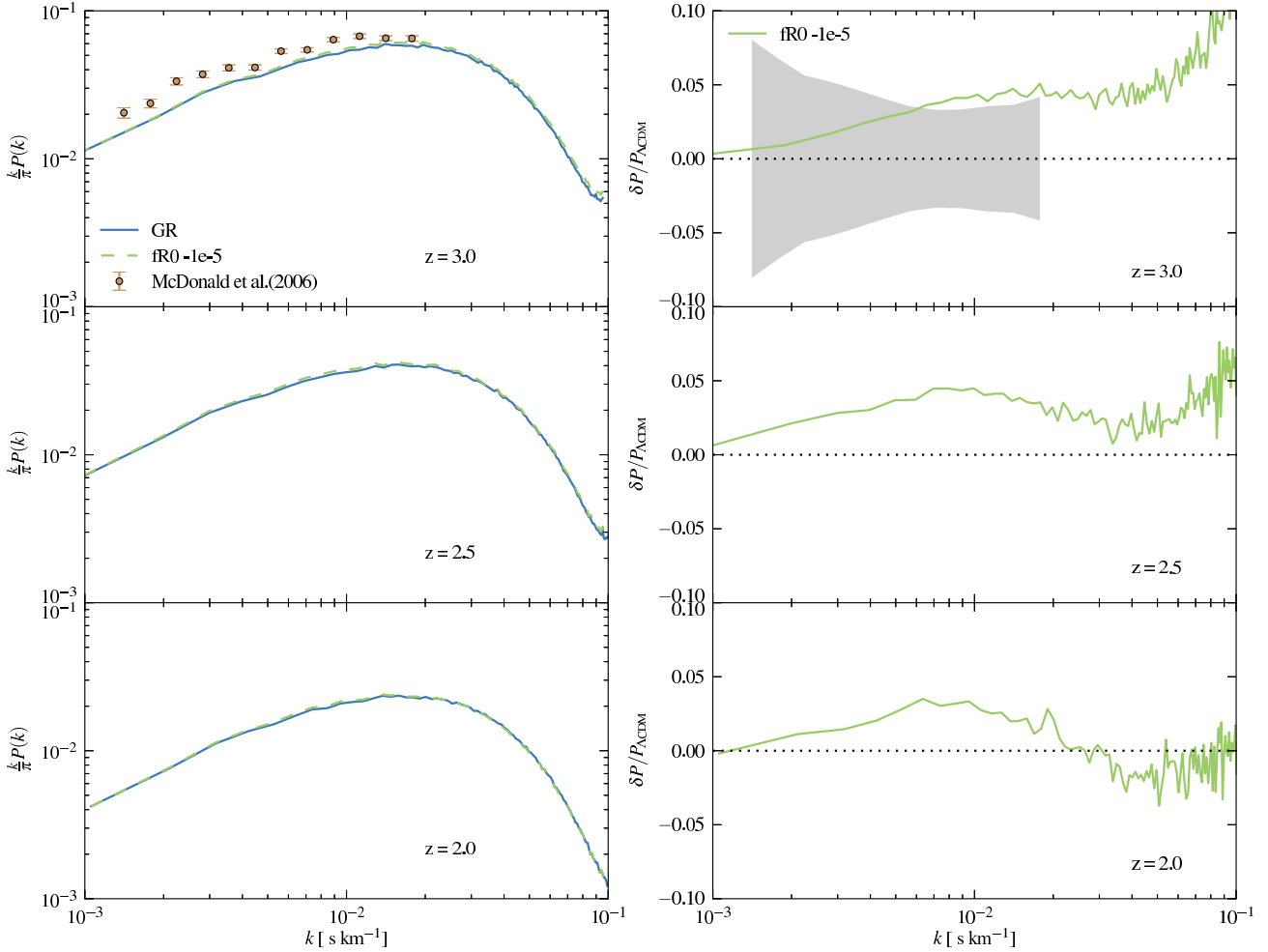}}
  \caption{\textit{Left panel:} Flux power spectra for $f(R)$-gravity
    and $\Lambda$CDM obtained from the $60\,h^{-1}\text{Mpc}$ simulation
    boxes at different redshifts. The dots with errorbars show the
    results of \protect\cite{mcdonald2006}. \textit{Right panel:}
    relative difference in the flux power spectra shown on the left
    hand side. The shaded area represents the relative errors of the
    \protect\cite{mcdonald2006} results at $z=3$. The mean
    transmission is tuned to the values of \protect\cite{kim2007} for
    both panels.}
\label{fig:fluxpower}
\end{figure*}

\begin{figure*}
  \centerline{\includegraphics[width=\linewidth]{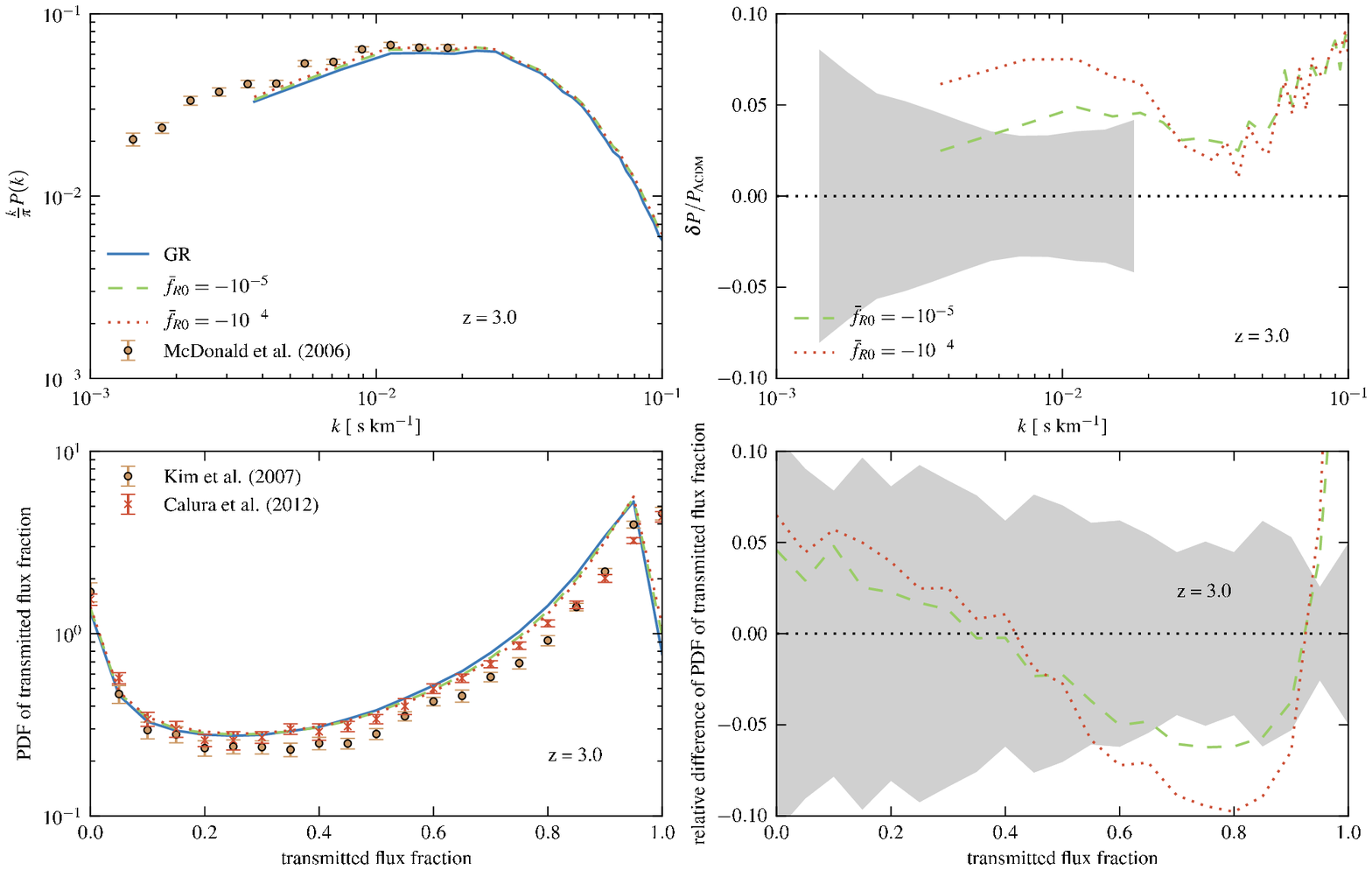}}
  \caption{\textit{Top panels:} Same as Figure~\ref{fig:fluxpower}.
    \textit{Bottom panels:} same as Figure~\ref{fig:fluxpdf}. In
    contrast to the previous figures the results are obtained from the
    $15\,h^{-1}\text{Mpc}$ simulation boxes and for both $|\bar{f}_{R0}| = 10^{-4}$ and $10^{-5}$.}
\label{fig:small_box}
\end{figure*}

Previous simulation works on chameleon-type $f(R)$-gravity mainly
employed collisionless simulations, analyzing, e.g., halo mass functions
\citep[e.g.][]{schmidt2009,ferraro2011,zhao2011b,lihu2011}, matter
power spectra
\citep[e.g.][]{oyaizu2008,li2012,li2013,puchwein2013,llinares2013},
density profiles \citep{lombriser2012, moran2014}, cluster
concentrations \citep{lombriser2012b}, the integrated Sachs-Wolfe
effect \citep{cai2013}, redshift space distortions
\citep{jennings2012}, velocity dispersions \citep{schmidt2010,
  lombriser2012b, lam2012, arnold2014}, the impact of screening on
the fifth force in galaxy clusters \citep{moran2014} 
and galaxy populations which were followed with a semi-analytical model \citep{fontanot2013}. Using
hydrodynamical simulations instead, we studied in previous work
\citep{arnold2014} the temperature of the intracluster medium and
different mass measures of galaxy clusters.

In this work, we will for the first time extend the analysis of
hydrodynamical $f(R)$ modified gravity simulations to the statistical
properties of the Lyman-$\alpha$ forest. Employing the modified
gravity simulation code \textsc{mg-gadget} we carry out simulations to
redshift $z = 2$, given that most Lyman-$\alpha$ data lies at redshifts
$z\sim 2-4$. Creating synthetic Lyman-$\alpha$ absorption spectra
from the simulation outputs, we present an analysis of flux PDFs, flux
power spectra, line shape statistics, as well as of the matter power
spectrum for both $f(R)$-gravity and an ordinary $\Lambda$CDM model.
We particularly focus on the relative differences between these two
cosmogonies and compare our results to observations.

This paper is structured as follows. In Section~2, we give a brief
summary of the particular parameterization of $f(R)$ we adopt, whereas
Section~3 describes the simulation set we have carried out. In
Section~4, we present our main results. Finally, we give a summary and
our conclusions in Section~5.

\section{$\MakeLowercase{f}(R)$-gravity}
\label{sec:fR_gravity}

$f(R)$ gravity models can explain the late time accelerated expansion
of the Universe without a cosmological constant. Using the framework
of Einstein's general relativity (GR), they add a scalar function
$f(R)$ to the action: \begin{align}
 S=\int {\rm d}^4x \sqrt{-g} \left[ \frac{R+f(R)}{16\pi G} +\mathcal{L}_m \right],\label{action}
\end{align}
where $\mathcal{L}_m$ is the matter density Lagrangian, $G$ is the gravitational constant and $g$ is the determinant of the metric. 
As in GR, the field equations are obtained by variation with respect to the metric, leading to the so called  \textit{modified Einstein's equations} \citep{buchdahl1970}:
\begin{align}
G_{\mu\nu} + f_R R_{\mu\nu}-\left( \frac{f}{2}-\Box f_R\right) g_{\mu\nu} - \nabla_\mu \nabla_\nu f_R = 8\pi G T_{\mu\nu} \label{Eequn}.
\end{align}
$f_R = \frac{\text{d}f(R)}{\text{d}R}$ denotes the derivative of the
scalar function $f(R)$ with respect to the Ricci scalar.

In order to simplify the above equation one can use the quasi-static
approximation and neglect time derivatives if only scales much smaller
than the horizon are considered \citep{oyaizu2008, noller2013}. In
addition, the calculations can be restricted to models with
$|f_R|\ll1$, as only these are compatible with observations, giving
(\citealt{oyaizu2008}, also see the Appendix of \citealt{arnold2014}
for a detailed derivation): \begin{align}
 \nabla^2 f_R = \frac{1}{3} (\delta R - 8\pi G\delta\rho). \label{nablafR}
\end{align}
$\delta R$ and $\delta\rho$ denote the perturbations to the
Ricci-scalar and the matter density with respect to the mean
background values at a given time $a$.

To complete the set of differential equations for $f(R)$-gravity in
the Newtonian limit, a modified Poisson equation for the gravitational
potential can be derived from equation~(\ref{Eequn})
(\citealt{husa2007}, see \citealt{arnold2014} for a more detailed
derivation): \begin{align}
 \nabla^2 \Phi = \frac{16\pi G}{3}\delta\rho - \frac{1}{6} \delta R.\label{poisson}
\end{align}
To simulate the evolution of matter in the Universe in $f(R)$-gravity,
equations (\ref{nablafR}) and (\ref{poisson}) have to be solved. But first,
the function $f(R)$ needs to be specified. It should be chosen such
that it meets the following requirements: First of all, the model
should reproduce the well constrained expansion history of a
$\Lambda$CDM universe. And second, the $f(R)$ modifications of
ordinary gravity should be screened in high density regions, because
GR appears valid to high accuracy in our local environment.

A model which is exactly designed to meet these requirements was
proposed by \cite{husa2007}: \begin{align}
 f(R) = -m^2\frac{c_1\left(\frac{R}{m^2}\right)^n}{c_2\left(\frac{R}{m^2}\right)^n +1},
\end{align}
where $m^2 \equiv H_0^2\Omega_m$. It produces a chameleon mechanism
which ensures that high curvature regions like our local environment
are screened from $f(R)$ effects and thus experience the same forces
as in GR.

The model reproduces the expansion history of $\Lambda$CDM if one
chooses the parameters $c_1$ and $c_2$ according to \citep{husa2007}:
\begin{align}
  \frac{c_1}{c_2}=6\frac{\Omega_\Lambda}{\Omega_m} && \text{and} &&
  c_2\left(\frac{R}{m^2} \right)^n \gg 1. \label{eq:c1_c2_lambda}
\end{align} $n$ will be set to $1$ for all simulations in this work.
The remaining free parameter can be expressed in a more convenient
way as follows. For a given choice of parameters, the derivative of
$f(R)$ with respect to $R$ reads: \begin{align}
  f_R=-n\frac{c_1\left(\frac{R}{m^2}\right)^{n-1}}{\left[c_2\left(\frac{R}{m^2}\right)^n+1\right]^2}\approx-n\frac{c_1}{c_2^2}\left(\frac{m^2}{R}\right)^{n+1},\label{fR}
\end{align} again using $c_2\left(\frac{R}{m^2} \right)^n \gg 1$ for
the second equality. Fixing the metric to a
Friedman-Lemaitre-Robertson-Walker universe and assuming a flat
$\Lambda$CDM expansion history, the background curvature is:
\begin{align}
  \bar{R}=3m^2\left[ a^{-3} + 4\frac{\Omega_\Lambda}{\Omega_m} \right].\label{R_bar}
\end{align}
With equation (\ref{R_bar}) one can now express the free parameter in
terms of the background value of $f_R$ at $a=1$, $\bar{f}_{R0}$.
Together with equations (\ref{eq:c1_c2_lambda}) and (\ref{fR}),
$\Omega_m$, $\Omega_\Lambda$, $H_0$ and $n$, $\bar{f}_{R0}$ fully
constrains the model and allows solving of the field equations.

\begin{figure}
  \centerline{\includegraphics[width=\linewidth]{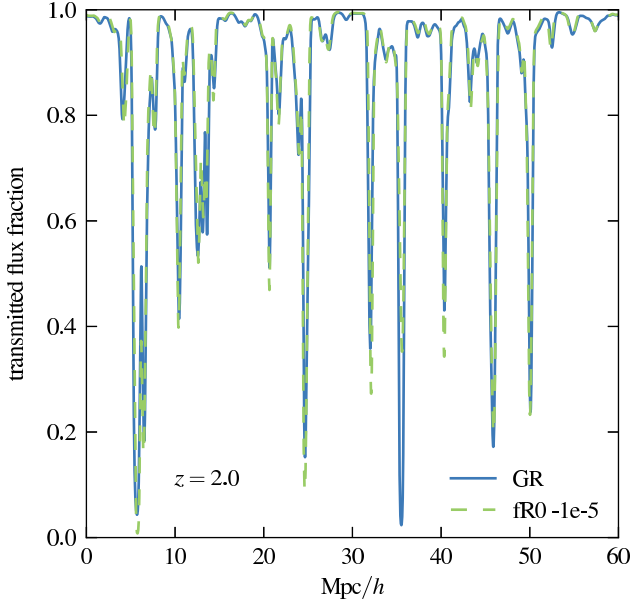}}
  \caption{Transmitted flux fraction along an arbitrarily selected
    line of sight at $z = 2$ for GR and $|\bar{f}_{R0}| = 10^{-5}$.}
  \label{fig:los_z0} 
\end{figure}

\section{Simulations and methods}
\label{sec:simulation_code}
Modeling the statistical properties of the Lyman-$\alpha$ forest
requires a cosmological simulation code which is capable of accounting
for a variety of gas physics, including photoheating, radiative
cooling and star formation. For $f(R)$-gravity, the fifth force influence has to be
computed in addition. In this work, we use the modified gravity
simulation code \textit{Modified-Gravity}-\textsc{gadget}
(\textsc{mg-gadget}, \citealt{puchwein2013}), which is an extension of
\textsc{p-gadget3}, which in turn is an advanced version of
\textsc{gadget2} \citep{springel2005c}. Featuring the gas-physics
modules of \textsc{p-gadget3} as well as a modified gravity solver,
\textsc{mg-gadget} offers the possibility to analyze the
Lyman-$\alpha$ forest in the \cite{husa2007} model of $f(R)$-gravity,
including nonlinear effects caused by the chameleon mechanism.

\textsc{mg-gadget} solves the $f(R)$-equations as follows (for full
details of the code functionality, see \citealt{puchwein2013}): First,
equation~(\ref{nablafR}) is solved using an iterative
Newton-Gauss-Seidel relaxation scheme. The iterations are carried out
on an adaptively refining mesh so that an increased resolution is
available in high density regions without causing a great loss of
performance. Solving for $f_R$ directly could lead to positive values
for $f(R)$ due to the finite iteration step-size. Because this would
immediately stop the simulations, the code solves for $u \equiv \ln
(f_R /\bar{f}_R(a))$ instead (following \citealt{oyaizu2008}), and then calculates $f$ from $u$. This
iterative scheme is well suited for the highly nonlinear behavior of
$f_R$.

\begin{figure}
  \centerline{\includegraphics[width=\linewidth]{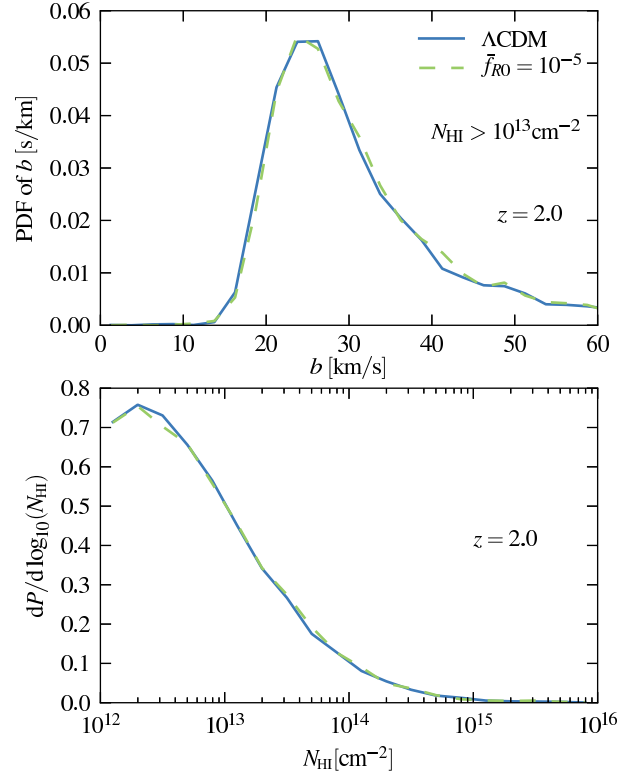}}
  \caption{\textit{Upper panel:} PDF of the linewidths of the Voigt
    profile fits to the Lyman-$\alpha$ absorption lines in the
    synthetic spectra for all lines with neutral hydrogen column
    density $N_{\rm HI} > 10^{13} \rm{cm}^{-2}$ at $z=2$.
    \textit{Lower panel:} normalized PDF of the column density,
    considering all lines with $N_{\rm HI} > 10^{12} \rm{cm}^{-2}$ at the
    same redshift. The spectra for both panels were tuned to the mean
    transmission of \protect\cite{becker2013} and were calculated from
    the $60\,h^{-1}\rm{Mpc}$ simulations.}
\label{fig:linewidth}
\end{figure}

As soon as the value for $f_R$ is known, one can rewrite the modified Poisson equation in the form
\begin{align}
 \nabla^2 \Phi = 4 \pi G (\delta\rho + \delta\rho_{\rm eff}),\label{poisson_mod}
\end{align}
where 
\begin{align}
 \delta\rho_{\rm eff} = \frac{1}{3}\delta\rho - \frac{1}{24\pi G}\delta R,\label{delta_rho_eff}
\end{align}
is an effective density perturbation. Calculating $\delta R$ through
equation~(\ref{fR}), the effective mass density is obtained. Adding
the effective density to the real mass density, the standard TreePM
gravity solver of \textsc{p-gadget3} can be used to solve
equation~(\ref{poisson_mod}) and calculate the gravitational forces on
the individual particles, including $f(R)$ effects. Note that all non-linearities of the model
are already encoded in $\delta\rho_{\rm eff}$ during this step. The hydrodynamical
forces are calculated using the smoothed particle hydrodynamics (SPH)
method implemented in \textsc{p-gadget3} \citep{Springel2002}.

For a reliable analysis of $f(R)$'s impact on the Lyman-$\alpha$
forest it is necessary to run simulations in both modified gravity and
$\Lambda$CDM using identical initial conditions. We do so by carrying
out hydrodynamical simulations using $2 \times 512^3$ particles
($512^3$ each gas and dark matter) in a $60\,h^{-1}\text{Mpc}$
box for $|\bar{f}_{R0}| = 10^{-5}$ and GR. To explore the parameter
space at a coarse level, we also did a set of smaller box simulations
at the same mass resolution, but using $2 \times 128^3$ particles in a
$15\,h^{-1}\text{Mpc}$ box for $|\bar{f}_{R0}| = 10^{-4}$, $10^{-5}$
and GR. Our set of cosmological parameters is $\Omega_m = 0.305,
\Omega_\Lambda = 0.695, \Omega_b = 0.048$ and $H_0 = 0.679$,
consistent with CMB constraints.

To extract the statistical properties of the Lyman-$\alpha$ forest,
synthetic absorption spectra were calculated at different redshifts.
Using the output of the hydrodynamical simulations, we randomly select
$5000$ lines-of-sight (LOS), each intersecting the simulation box
parallel to one of the three coordinate axes. Dividing each line into
$2048$ pixels, the density of the neutral hydrogen, the gas
temperature and the neutral gas weighted velocity fields of the gas
are computed along the lines of sight. For this computation, we
consider all SPH particles whose smoothing length is intersected by
the sight-line. With the bulk flow velocities and temperatures along
the line of sight in hand, we then account for kinematic Doppler
shifts and thermal broadening of the absorption lines, which are
themselves calculated from the neutral hydrogen density. As final
output, the code creates for each LOS a file with optical depth $\tau$
as a function of distance along the LOS. This output can be converted
into a transmitted flux $F=\text{e}^{-\tau}$. 

\begin{figure}
  \centerline{\includegraphics[width=\linewidth]{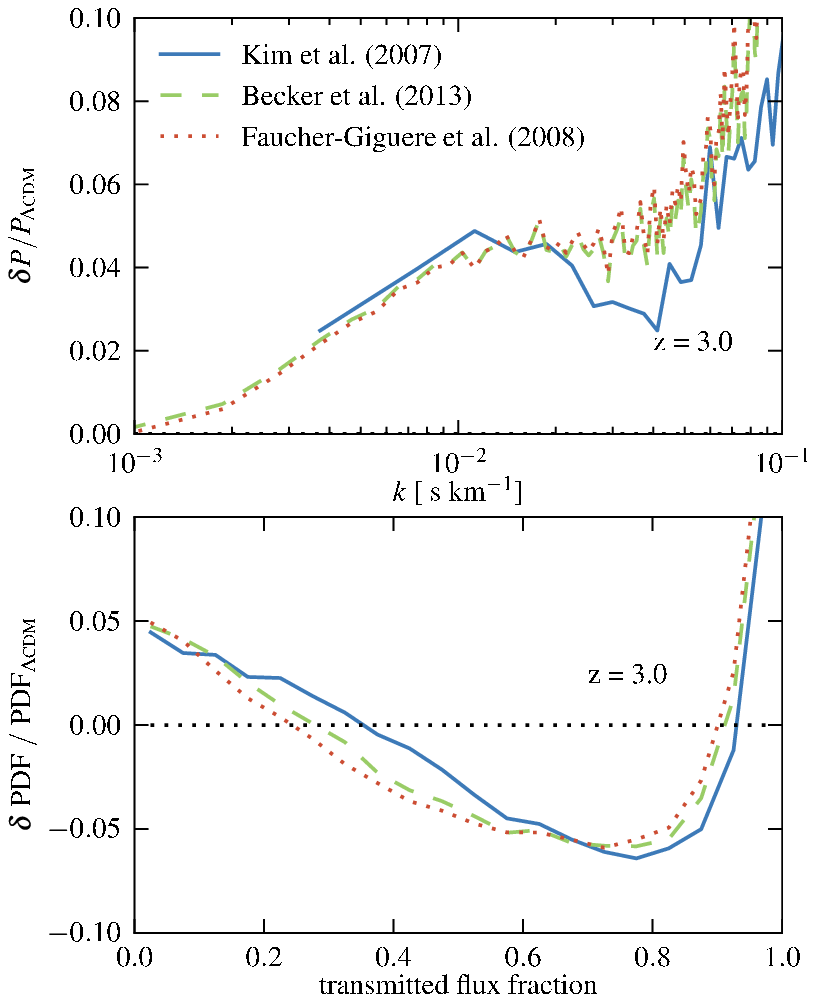}}
  \caption{\textit{Upper panel:} relative difference between $f(R)$
    and $\Lambda$CDM in the Lyman-$\alpha$ flux power spectrum;
    \textit{lower panel:} relative difference between modified gravity
    and GR in the PDF of the transmitted flux fraction. For both
    plots, the values of the mean transmitted flux were tuned either
    to \protect\citet[][\textit{blue solid line}]{kim2007},
    \protect\citet[][\textit{green dashed line}]{becker2013}, or
    \protect\citet[][\textit{brown dotted line}]{faucher2008},
    respectively. The results refer to $z=3$.}
\label{fig:compare_reldiff}
\end{figure}

In the simulations, \textsc{mg-gadget} uses a tabulated UV-background\footnote{
Taken from \citet{haardt2012}, however with the He\,\textsc{ii} photoheating rate boosted by a factor $1.7$ for $2.2 < z < 3.4$.
This slight modification results in a better agreement with observational constraints \citep{becker2011} on the temperature of the intergalactic medium.}
which allows the calculation of gas temperatures and ionization
fractions assuming ionization equilibrium. The results
might however not fit the mean Lyman-$\alpha$ flux transmission
actually seen in the observational data. In order to compare the mock
spectra to observations more faithfully, it is therefore necessary to
tune the mean transmitted flux to the corresponding value obtained in
the observations. We perform this rescaling of the optical depths, which is a standard
procedure in theoretical studies of the Lyman-$\alpha$ forest, in our post-processing of the LOS data.

\section{Results}
\label{sec:results}

In analyzing the synthetic Lyman-$\alpha$ absorption spectra, we
first consider the PDF of the transmitted flux fraction in the
$60\,h^{-1}\text{Mpc}$ box simulations for both $|\bar{f}_{R0}| =
10^{-5}$ and $\Lambda$CDM. Figure \ref{fig:fluxpdf} shows these PDFs
at redshifts $z=2$, $2.5$ and $3$ (\textit{left hand panels}), as well
as the relative differences between the two cosmological models
(\textit{right hand panels}). The mean transmitted flux fraction is
tuned in this plot to the observational values of \cite{kim2007}. We
also show this data in the panels on the left hand side and its
relative errors on the right hand side, for reference. 
In addition, the results of \cite{calura2012} are shown for redshift $z = 3$.
In both observations metal lines were excised, in the latter also Lyman-limit systems (LLS). 
The mean transmission measured by \cite{calura2012} is consistent with the one of \cite{kim2007}. We can thus compare to the same theoretical prediction.

Comparing the absolute values, it is obvious that, despite the tuning
to the same  effective $\bar{\tau}$, the simulation results do not match the
observations particularly well. 
Especially at redshift $z=3$, the gap
between the simulation results of both $f(R)$ gravity and GR and the
observational values of \cite{kim2007}  is much larger than the errorbars.
At intermediate fluxes, the \cite{calura2012} results are much closer to the simulations. 
Nevertheless, the differences at large transmitted flux fractions clearly exceed the $3 \sigma$ observational error.
Considering the panels for redshift $z=2$ and $2.5$, the discrepancies
between simulations and the observational data are smaller, but in
certain regimes still larger than $3 \sigma$. These differences might
have their origin in the still uncertain systematic errors of
observations (like, e.g., in the continuum placement) and simulations, in an underestimate of the statistical errors \citep{rollinde2013} or in an unaccounted heating of 
the very low-density intergalactic medium \citep{bolton2008,viel2009} by radiative transfer \citep{mcquinn2009,compostella2013}
and non-ionization-equilibrium \citep{puchwein2014} effects or, as recently suggested, by TeV blazars \citep{broderick2012,puchwein2012}.

\begin{figure*}
  \centerline{\includegraphics[width=\linewidth]{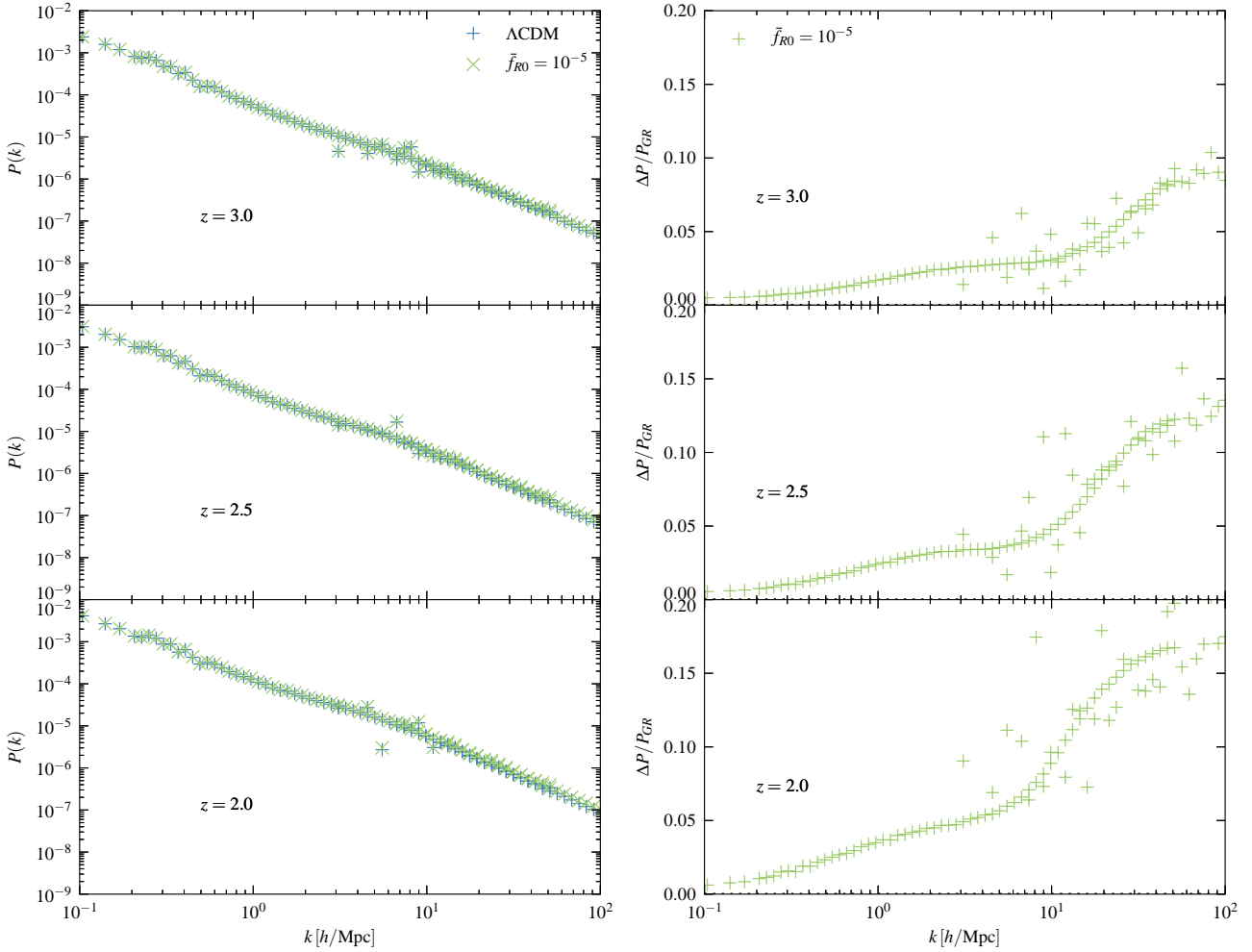}}
  \caption{\textit{Left Panel:} matter power spectrum for $\Lambda$CDM
    and $|\bar{f}_{R0}| = 10^{-5}$ at three different redshifts.
    \textit{Right panel:} Corresponding relative difference in the
    power spectra $(P_{f(R)}-P_{\rm GR})/P_{\rm GR}$ between
    $f(R)$ and GR.}
\label{fig:matterpower}
\end{figure*}

The difference between $|\bar{f}_{R0}| = 10^{-5}$ and $\Lambda$CDM is
much smaller than the
difference between the simulation results and the  
observed values and somewhat smaller than the errorbars of the observations. This is even more obvious in the right
hand panels of Figure~\ref{fig:fluxpdf}. Comparing the relative
difference in the flux PDF between $f(R)$ and GR, it turns out that
the differences between the models are mostly within the observational errors for individual flux bins
for all considered redshifts. The overall deviation over many flux bins and redshifts could, however,
still be statistically significant if systematic effects were better understood.
 Given the current uncertainties in Lyman-$\alpha$ forest studies and the already very tight
constraints on the \cite{husa2007} model of $f(R)$ gravity from other
observables, it will therefore be hard to get competitive constrains on $|\bar{f}_{R0}|$
using the flux PDF of the Lyman-$\alpha$ forest.


We arrive at a similar conclusion for the Lyman-$\alpha$ flux
power-spectra obtained for the large simulation box:
Figure~\ref{fig:fluxpower} shows their absolute value for both $f(R)$
gravity and a $\Lambda$CDM cosmology as well as the relative
difference between these models. The theoretical results from our
synthetic spectra are again tuned to the mean transmission of
\cite{kim2007}. For $z = 3$, the simulation values are compared to the
observational results of \cite{mcdonald2006}, with the gray shaded
area in the relative difference plot indicating their quoted errors.

As for the flux PDF, the discrepancy between simulations and
observations at redshift $z=3$ is quite large, in particular much
larger than the errors given in \cite{mcdonald2006}. This might have
its origin in systematic uncertainties which are not considered by
the errorbars. Again, the difference between the two gravitational
models is tiny, compared to the difference between observational data
and the results from the simulations. Because the difference to GR is
again smaller than or comparable to the errorbars of the shown observations, we need to
conclude that the Lyman-$\alpha$ flux power-spectrum is only mildly
affected by $f(R)$-gravity.

This is also particularly evident in the relative difference plots at the
right hand side. The difference in the flux power-spectrum between
$|\bar{f}_{R0}| = 10^{-5}$ and GR is about $5\%$ at maximum,
considering redshifts $z = 2$, $2.5$ and $3$. Normalizing the results
of \cite{mcdonald2006} to our $\Lambda$CDM outcome, it is obvious that
the relative difference between the $f(R)$ simulation results and the
fiducial model is consistent with the relative errors quoted for the individual $k$ bins. 
The overall deviation over many bins and redshifts may be statistically significant. However, systematic effects would
need to be better understood to obtain interesting constrains on $\bar{f}_{R0}$ based on such observations.

To test if it is at all possible to constrain $f(R)$ gravity using the
Lyman-$\alpha$ forest, we also run a set of simulations with smaller
box size at equal mass and spatial resolution, for $|\bar{f}_{R0}| =
10^{-4}, 10^{-5}$ and GR. The power spectra and flux PDFs at redshift $z =
3$ are shown in Figure~\ref{fig:small_box}. For both power spectra and
the PDFs, the results for GR and $|\bar{f}_{R0}| = 10^{-5}$, as well
as their relative differences, are compatible with the values from the
bigger simulation box shown in Figures~\ref{fig:fluxpdf} and
\ref{fig:fluxpower}. One can therefore conclude that the smaller box
runs are sufficient for an analysis over the shown range of values. In
the $|\bar{f}_{R0}| = 10^{-4}$ simulations, the flux power spectrum
does not fit the observed values of \cite{mcdonald2006} despite tuning
the mean transmission. As the absolute value of the Lyman-$\alpha$
power spectrum is not known with great accuracy, one should not
overestimate the, compared to the over gravitational models, smaller
difference of the $|\bar{f}_{R0}| = 10^{-4}$ curve to the
observations. Again, normalizing the GR results to the observations,
one can compare the observational errors to the differences between
$f(R)$ and a $\Lambda$CDM universe. At intermediate scales the
difference between $|\bar{f}_{R0}| = 10^{-4}$ and GR is larger than
for $|\bar{f}_{R0}| = 10^{-5}$. Nevertheless it does not exceed the $2
\sigma$ relative error of the observations for individual $k$ bins. Given that 
$|\bar{f}_{R0}| = 10^{-4}$ appears already clearly ruled out by other
methods \citep{lombriser2012b, lombriser2012, smith2009, schmidt2009,
  dosset2014}, it does not seem that current Lyman-$\alpha$ data can add much new information here.

Figure~\ref{fig:small_box} also shows the PDF of the transmitted flux
for the three different gravity models. As for the power spectrum, the
$|\bar{f}_{R0}| = 10^{-4}$ values do not fit the data much better than
the other models at $z = 3$. Comparing relative errors to the
difference between the models, we see that the difference
between $|\bar{f}_{R0}| = 10^{-4}$ and GR lies within the $1
\sigma$-error region for almost all values. Only between a transmitted
flux fraction of $0.6$ and $0.9$ the difference is larger than $1
\sigma$ and reaches about $2 \sigma$ at maximum. The flux PDF does
therefore also not seem to be very competitive with current
observational data compared to other methods to constrain $\bar{f}_{R0}$.

In comparison to other uncertainties in the cosmological model, the
impact of $f(R)$ gravity on Lyman-$\alpha$ flux power spectra or PDFs
is fairly small, even if one considers quite extreme and already
excluded values for $|\bar{f}_{R0}|$.

Figure~\ref{fig:los_z0} illustrates
how small the modified gravity effect on the Lyman-$\alpha$ forest is. It displays the transmitted flux fraction along
an arbitrarily selected line of sight for $f(R)$ and $\Lambda$CDM as a
function of distance along this line. The positions of the absorption
lines are the same for both models, as identical initial
conditions have been used in both simulations. While there appear slight differences in the transmitted
flux fractions for the individual absorption lines, no general pattern
can be identified from a visual inspection.

Taking a more systematic approach, we used the code AutoVP
\citep{autovp} to fit Voigt-profiles to the absorption lines of the
synthetic spectra. The PDF of the linewidth of the fitted
Voigt-profiles is shown in the upper panel of
Figure~\ref{fig:linewidth} for all lines with a neutral hydrogen
column density $N_{\rm HI} > 10^{13} \rm{cm}^{-2}$ at redshift $z=2$.
As the lines almost perfectly overlap each other, it is perhaps not
surprising that there is no significant difference in the linewidth
distributions between $f(R)$ gravity and a $\Lambda$CDM universe. The
lower panel of the figure displays the normalized column density PDF
of the absorption lines. As for the linewidth, the difference between
the curves for $|\bar{f}_{R0}| = 10^{-5}$ and the $\Lambda$CDM model
is negligible.

The absolute values of the statistical Lyman-$\alpha$ measures depend
on the observational value the mean transmitted flux is tuned to. To
justify our previous analysis, we briefly show that the {\rm relative
  differences} do not depend strongly on the actual value that is adopted.
Figure~\ref{fig:compare_reldiff} shows the relative difference in flux
PDF and power spectrum between $|\bar{f}_R| = 10^{-5}$ and GR. Each
line in the plot is tuned to a different mean $\tau$, representing the
observational data of \cite{becker2013}, \cite{kim2007} and
\cite{faucher2008}. The figure shows that neither the
relative difference in the power spectra nor the flux PDF do strongly
depend on the choice of the tuning value for $z=3$. As our analysis
confirms, this does also hold for $z=2$ and $2.5$. One can therefore
conclude that the relative differences can be explored safely despite
the fact that the absolute values depend on the actual value used for
the mean transmission.

To complement our analysis of the Lyman-$\alpha$ forest in the
simulations, we also analyzed the total matter power spectra at
different redshifts. Figure~\ref{fig:matterpower} shows these spectra
for $|\bar{f}_R| = 10^{-5}$ and $\Lambda$CDM scenarios at redshifts
$z=2$, $2.5$ and $3$, as well as the relative difference between the
models. Comparing the power spectrum enhancement to previous works,
our results at redshift $z=2$ are in good agreement with those of
\cite{li2013}. The evolution of the enhancement with time in our
simulations is consistent with previous works, too \citep{li2013,
  winther2014}. This confirms that our $f(R)$ simulations feature an
impact of modified gravity at the expected level, even though the
effects on the Lyman-$\alpha$ forest properties are weak.

Comparing matter and flux power spectra at the same scale, the relative differences between $f(R)$ and GR
are of similar size. The matter power spectra exhibit larger differences at smaller scales. There the Lyman-$\alpha$ flux power spectrum
becomes, however, degenerate with uncertainties in the temperature of the intergalactic medium.
Other gas properties like the temperature of gas in collapsed objects 
typically show $f(R)$ effects of order
$30\%$ in the unscreened regime \citep{arnold2014}. This also illustrates that the impact of
$f(R)$ gravity on the Lyman-$\alpha$ forest is  rather small in
comparison. In order to constrain $f(R)$, it therefore appears more
promising to focus on other measures of structure growth rather than
the Lyman-$\alpha$ forest.

\section{Summary and Conclusions}
\label{sec:conclusions}

In this work, we analyzed the statistical properties of the
Lyman-$\alpha$ forest and the matter power spectra in hydrodynamical
cosmological simulations of $f(R)$ gravity. Our simulations employed
the \cite{husa2007} model and used the modified gravity simulation
code \textsc{mg-gadget}. For comparison, we also ran a set of
simulations for a $\Lambda$CDM universe.
Our main findings can be summarized as follows:
\begin{itemize}
\item The PDF of the transmitted Lyman-$\alpha$ flux fraction is only
  mildly affected by $f(R)$ gravity. The maximum relative difference between
  $|\bar{f}_{R0}| = 10^{-5}$ and GR is of order $7\%$. For
  $|\bar{f}_{R0}| = 10^{-4}$ the difference grows to at most $10\%$.
  The simulation results do not fit the observational data of
  \cite{kim2007} at all redshifts, regardless of the gravitational
  model. If the observations are normalized to the GR results from the
  simulations, the relative differences between the gravitational
  models do not exceed the $2\,\sigma$ relative error of the
  observations.
  For $z=3$, the \cite{calura2012} data matches the simulation results at intermediate transmitted flux fractions much better. 
  This highlights that the present observational data is relatively uncertain. At high transmissions there are also significant deviations.

\item For the flux power spectra we arrive at similar results. The
  relative difference between the models reaches at most $5\%$ for
  $|\bar{f}_{R0}| = 10^{-5}$ and about $10\%$ for $|\bar{f}_{R0}| =
  10^{-4}$. Again, the differences to GR for the stronger model are
  within the $2\,\sigma$ relative error of the observational data
  \citep{mcdonald2006}. Despite tuning the mean transmitted flux to
  the observational values, the power spectrum at $z=3$ does not
  accurately reproduce the observational data. 

\item There is no significant change in the shapes and abundances of
  absorption lines in $f(R)$ modified gravity: Both the column density
  and line
  width distribution functions based on Voigt profile fitting do not exhibit any systematic change.
 
\item The relative
  differences between $f(R)$ and GR in the flux PDF and power spectra do
  not depend significantly on the observed mean transmission value to which the simulated spectra are scaled. The tuning affects only the
  absolute values of these statistical Lyman-$\alpha$ measures. 
  
\item
  The matter power spectrum shows an enhancement in $f(R)$ gravity
  which grows with time. The amplitude of the effect and the relative
  difference to GR is consistent with previous works. 
  These relative changes in
  the matter power spectrum are of comparable magnitude as the changes in the Lyman-$\alpha$ forest flux power spectrum at the same scale. Note, however, that the 
  enhancement of the matter power spectrum continues to grow towards low redshift where it is no longer probed by the Lyman-$\alpha$ forest. Also, note
  that a much stronger influence of $f(R)$ has been found for other gas properties like the gas temperature in collapsed objects in the unscreened regime as has been reported in \citet{arnold2014}.

 \end{itemize}

 All in all we arrive at the conclusion that the impact of $f(R)$
 gravity on the Lyman-$\alpha$ forest is small. The relative differences in flux PDFs
 and flux power spectra are only of the order of $5\%$. Even for likely 
 excluded models, like $|\bar{f}_{R0}| = 10^{-4}$, the changes in the
 statistical Lyman-$\alpha$ forest properties do not exceed the
 relative errors of available observations in individual flux or wavenumber bins. 
 Using the full data over a range of redshifts a detection of modified gravity effects could probably be statistically significant due to its clearly defined signature.
 However, currently, systematic effects do not seem to be understood at the required level to get competitive constraints in practice. 
 One can therefore conclude that Lyman-$\alpha$ forest properties are of
 limited discriminative power to constrain $|\bar{f}_{R0}|$ at the moment. The
 remarkable robustness of the forest statistics has however also
 advantages. Given that the considered gravitational models
 have a negligible impact on the Lyman-$\alpha$ forest compared to
 other cosmological and astrophysical uncertainties, it may not be necessary to
 consider $|\bar{f}_{R0}|$ as an additional parameter in constraining
 cosmological parameters based on the
 Lyman-$\alpha$ forest.

\section{Acknowledgements}

C.A. and V.S. acknowledge support from the Deutsche
Forschungsgemeinschaft (DFG) through Transregio 33, ``The Dark
Universe''. E.P. would like to thank Martin Haehnelt for helpful discussions 
and acknowledges support by the FP7 ERC Advanced Grant Emergence-320596. 

\bibliographystyle{mn2efixed}
\bibliography{paper}

\end{document}